# A Prototype for a Controlled and Valid RDF Data Production Using SHACL

Elia Rizzetto[1], Arcangelo Massari[1], Ivan Heibi[1,2] and Silvio Peroni[1,2,*,†]

[1]*Department of Classical Philology and Italian Studies, University of Bologna*
[2]*Research Centre for Open Scholarly Metadata, Department of Classical Philology and Italian Studies, University of Bologna, Bologna, Italy*

**Abstract**
The paper introduces a tool prototype that combines SHACL's capabilities with ad-hoc validation functions to create a controlled and user-friendly form interface for producing valid RDF data. The proposed tool is developed within the context of the OpenCitations Data Model (OCDM) use case. The paper discusses the current status of the tool, outlines the future steps required for achieving full functionality, and explores the potential applications and benefits of the tool.

## 1. Introduction

As RDF datasets grow larger, more complex, and are getting highly used/integrated into several services, ensuring the quality and validity of such data is a crucial aspect[1]. RDF data validation plays a pivotal role in enhancing interoperability, facilitating data integration, and ensuring data consistency across different applications and domains. To consider RDF data valid, it has to follow some specific constraints. Using a validation schema, organizations and developers can detect errors/inconsistencies, leading to improved data quality and reliability.

Despite the presence of many standards for inference like RDF Schema and OWL, these technologies employ Open World and Non-Unique Name Assumptions which creates difficulties for validation purposes[2]. Using Shape Expressions language (ShEx) we can fulfill a role similar to that of Schema languages for XML but specifically applied to RDF graphs[3]. Shape Expressions (ShEx) defines a concise, formal, modeling and validation language for RDF structures[4].

Another valid alternative is the Shapes Constraint Language (SHACL)[5]. In SHACL, validation is based on shapes, which define particular constraints and specify which nodes in a graph should be validated against these constraints. A set of constraints can also be interpreted as a "schema", functioning as one of the primary descriptors of a graph dataset. A SHACL document stores a set of SHACL shapes also called shapes graph[6].

While SHACL has a more verbose and complex syntax compared to ShEx, on the other hand, it provides a rich set of constraints and validation schemas, that allows complex shape definitions and advanced rule-based validations[7]. Also, SHACL has wider adoption (a larger integration into different RDF processing tools, e.g., including validators, editors, and integration with popular RDF libraries), so it also benefits from bigger ecosystem support[8].

Following these considerations, we decided to adopt SHACL into the work presented in this paper. Here we propose a first tool prototype that combines the potentials of SHACL with other ad-hoc validation functions, to create a controlled user-friendly form interface for the production of valid RDF data, ready to be further integrated into a triplestore. The process is tested in the context of the OpenCitations Data Model (OCDM) use case[9]. We conclude this paper through the discussion of the current tool status, the next steps to be accomplished to make it fully functional, and the future perspectives and potential usages of the tool.

## 2. Prototype

The production and modification of semantic data can be a particularly challenging task for inexperienced users of the Semantic Web, as it requires a certain degree of familiarity with the RDF language. For this reason, this work proposes the use of a user-friendly web interface (HTML form) that allows users to submit their data in a more familiar and intuitive environment.

The prototype logic is conceptualized in Figure 1, the core of the tool is a software component that takes as input a SHACL-expressed schema and a series of ad-hoc validation functions. Then, following the definitions made in the two modules, a web interface is generated through which users can enter and modify data related to entities and their properties. Following user intervention and data submission, the data is validated in two subsequent phases:

- Validation against SHACL shapes: This validation ensures the validity of the data regarding aspects such as the mandatory presence of a specific RDF property for a given entity (or class of entities), specific data types for property values, a range of possible values for certain properties, a minimum and/or a maximum number of properties for certain nodes, etc. SHACL also allows for the verification of more complex relationships between entities in the data graph by applying conditional constraints expressed through the use of SPARQL queries.
- Validation against property ad-hoc validation functions: using specifically implemented validation functions, we further restrict what the user can add or modify, ensuring greater granularity in controlling aspects of the data that cannot be represented and controlled with SHACL shapes. For instance, to verify the availability and accessibility of the resources, by definig specific functions to perform API requests to external services. Furthermore, targeted programmatic solutions can filter correct values more precisely and be adaptable to the use case. For example, while SHACL allows the use of regular expressions to filter possible values for certain properties, custom validation functions could enable the management of more complex patterns, along with live validation, which would assist the user during the form-filling process even before the submission required for the validation of the remaining data.

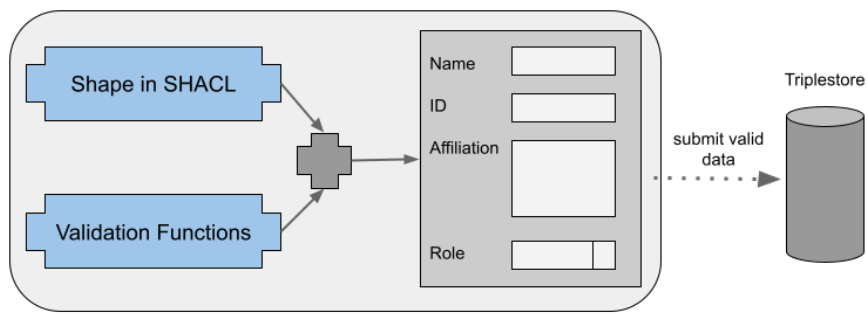

**Figure 1:** The prototype workflow, takes as input the SHACL shape and the validation functions to generate a web interface through which users can enter the attributes of a resource to be created and added to the triplestore.

Once the values to define a new resource are compiled using the web interface, the tool first checks whether the SHACL rules are respected and then it performs the additional validation functions for each property. Although the SHACL rules could be applied in the real-time compiling process, we might consider applying the rules of the validation functions only once, at the end of the compilation, i.e., once the submission is triggered. If the inputted values are valid and respect all the rules and constraints, the RDF data produced is ready to be submitted to the corresponding triplestore.

## 3. OpenCitations Data Model

The OpenCitations Data Model (OCDM) is based on a set of classes and properties that reflect the basic structures of the bibliographic domain, used to represent information about bibliographic resources and their related citations. It reuses entities defined in SPAR Ontologies to represent bibliographic entities (fabio:Expression), identifiers (datacite:Identifier), agent roles (pro:RoleInTime), responsible agents (foaf:Agent), and publication format details (fabio:Manifestation). These entities can be mapped using SHACL to define constraints that can then be used for the automatic generation of forms that enforce user adherence to these constraints.

For instance, let's consider a bibliographic resource, fabio:Expression. We can define a SHACL shape that represents the constraints of this resource as in Figure 2a. This schema provides a formal definition of the constraints that an entity of type fabio:Expression must adhere to. It indicates, for example, that the type of entity can be one of the listed ones (like fabio:Book, fabio:JournalArticle, etc.), and that the entity can have only one title, represented as a string (xsd:string).

Thanks to this formal definition, it is possible to automatically generate a data entry form that ensures compliance with the defined rules. For example, for the creation or modification of an entity of type fabio:Expression, a form generated based on this SHACL shape will have a "select" type input field for the selection of the resource type among those allowed, and a textual input field for the title entry (see Figure 2b).

In addition, a user could be also asked to insert a corresponding DOI in a text input box if

```
# BibliographicResource
schema:BibliographicResourceShape
    a sh:NodeShape ;
    sh:targetClass fabio:Expression ;
    rdfs:subClassOf schema:BibliographicEntityShape ;
    sh:property
    [
        sh:path rdf:type ;
        sh:in (fabio:ArchivalDocument
               fabio:Book
               fabio:BookChapter
               fabio:JournalArticle
               fabio:Thesis
               fabio:ProceedingsPaper) ;
        sh:minValue 1 ;
        sh:maxValue 2 ;
    ] ;
    sh:property
    [
        sh:path dcterms:title ;
        sh:datatype xsd:string ;
        sh:minValue 0 ;
        sh:maxValue 1 ;
    ] ;
.
```

(a) SHACL shape     (b) Web form

**Figure 2:** In (a) the SHACL shape of a bibliographic resource of type fabio:Expression within OCDM. It ensures that the type of an entity is one among the provided list (e.g., fabio:Book) and that the entity can have only one title, of string type (xsd:string); While (b) shows the form generated following the constraints defined using SHACL

the resource type is "Journal Article". In this case, on data submission, the ad-hoc validation functions are triggered to check the validity of the DOI value by calling the https://doi.org service.

## 4. Conclusions and Future Works

In this paper, we presented a prototype that currently is meant for users willing to create and add controlled/valid RDF data to a particular triplestore using a web form. Yet ideally we would like to generalize more this process through the inclusion of other common services used to interact with triplestore, such as API requests or SPARQL editors. In addition, a further feature to be considered is the bulk upload of the data to be added.

We consider the presented tool not as a stand-alone software, but rather as a plugin to be adopted in the future into larger systems. With this regard, two systems come to mind. One is CLEF (Crowdsourcing Linked Entities via web Form) [10], an agile LOD-native platform for collaborative data collection, peer-review, and publication. The second one is ResearchSpace, an open source platform designed to help establish a community of researchers, where their underlying activities are framed by data sharing, active engagement in formal arguments, and semantic publishing[11].

We are currently working toward the integration of this tool inside CLEF, and are willing to perform the first tests using the OpenCitations use case.